\documentclass[osajnl,a4paper,twocolumn]{revtex4}  

\usepackage{graphicx}

\begin{document}

\title{Synthetic-aperture experiment in the visible with on-axis digital heterodyne holography}

\author{F. Le clerc and M. Gross}
\address{
Laboratoire Kastler-Brossel, UMR 8552 CNRS, Ecole Normale Sup\'{e}rieure, Universit\'{e} Pierre et Marie Curie, 24 rue Lhomond 75231 Paris cedex 05 France\\
}
\author{L. Collot}
\address{
Thomson CSF Optronique, Rue Guynemer B.P. 55, 78 283 Guyancourt, France\\
}


\begin{abstract}
We have developed a new on-axis digital holographic technique, heterodyne holography. The resolution of
this technique is limited mainly by the amount of data recorded on two-dimensional photodetectors, i.e., the
number of pixels and their size. We demonstrate that it is possible to increase the resolution linearly with the amount of recorded data by aperture synthesis as done in the radar technique but with an optical holographic
field.
\end{abstract}



\maketitle

A synthetic aperture is used in a technique for signal
processing that combines signals acquired by
a moving detector into a unique signal field map
that permits higher-resolution image reconstruction.
This technique, which has been known for 30 years
in the visible \cite{reynolds1970imaging} and the near IR \cite{dereniak1973application}, is used extensively
for synthetic-aperture radar \cite{cutrona1990synthetic} and synthetic-aperture
sonar \cite{special_issue_1992}. More recently, synthetic-aperture telescopes,
in which several small telescopes collect optical fields
that are combined to yield higher-resolution images,
have been developed \cite{baldwin1996first}. For all these syntheticaperture
techniques the ultimate angular resolution
$\lambda/D'$ corresponds to the equivalent aperture size $D'$,
i.e., to the detector displacement or to the telescopes'
maximum relative distance.

Digital holography permits the complex optical field
to be measured directly and images to be obtained
by calculation of the field in the plane of the object,
but, for both off-axis \cite{schnars1994direct} and on-axis \cite{decker1978electronic,zhang1998three} holography, the
finite number of recorded pixels limits the resolution
(which is nevertheless better in the on-axis case). To
overcome this limitation it is natural to construct a
holographic synthetic aperture by recording several
different holograms of the same object and reconstructing
the image from all of them. Here we use our
heterodyne holography setup \cite{le2000numerical} to illustrate this idea.
Using holograms that correspond to various positions
of the spatial filter, which selects the photons that
fulfill the sampling condition, we performed an aperture
synthesis in Fourier space. For a short-distance
object, we demonstrate the synthetic-aperture effect
by overcoming the limit on pixel-size resolution, which
is that of one hologram \cite{le2000numerical}.

\begin{figure}
\begin{center}
  \includegraphics[width=8 cm]{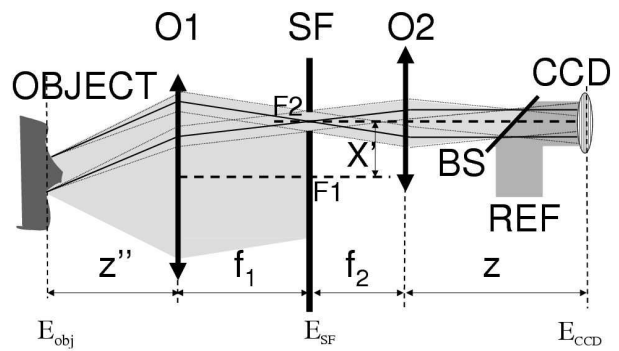}
  \caption{Spatial filter system composed of lenses O1 and
O2, rectangular spatial filter SF, beam splitter BS and a
CCD camera.}\label{Fig_1}
\end{center}
\end{figure}

Figure \ref{Fig_1} is a schematic of our experiment. The
signal field, at wavelength l, which is scattered by
a coherently illuminated object, passes through a
spatial filter system, that comprises objectives O1
($f_1=50$ mm) and O2 ($f_2 = 25$ mm), with rectangular
aperture spatial filter (SF) in their common focal
plane. O1 transforms field $E_{obj}$ in the plane of the
object into its $k$-space components $E_{SF}$ in the SF
plane, whereas O2 backtransforms $E_{SF}$ into $E_{CCD}$
on the CCD. O2, SF, and the CCD, which can be
moved, are kept aligned, and the field, when it reaches
the CCD, propagates nearly parallel to the reference
beam (REF). In Fig. \ref{Fig_1} and in what follows, the
$x''y''$ (object plane) coordinates refer to the O1 fixed
axis; $xy$ (CCD plane) and $x'y'$ (SF plane) coordinates
refer to the SF, CCD, and O2 moving axis, which
is shifted by $X', Y'$ with respect to O1. Double,
single, and no primes are used for the object, SF,
and CCD planes, respectively. The resolution on the
reconstructed object is $r'' = \lambda f_1/D'$, where $D'$ is the
size of the SF plane region where $E_{SF}$ is measured.
To avoid aliasing in the reconstructed images, $D'$
must fulfill the sampling condition $D' < D_{SF}= \lambda f_2/d$,
which corresponds to pixel size $d$. In the best case,
$D'= D_{SF}$, resolution $r''= d f_1/f_2$ is equal to the size
of the CCD pixels enlarged by the O1–O2 expansion
factor. We recorded holograms for several positions
of the SF, CCD, and O2 moving axis and determined
$E_{SF}$ over a displacement of the SF Fourier plane of
size $D'=  G.D_{SF}$ ($G \gg 1$).

For a given moving axis position ($X'_i,Y'_j$), the field
$(E_{SF})_{i,j}$ that is selected by SF is related to the measured
CCD field $(E_{CCD})_{i,j}$ by
\begin{equation}\label{Eq_1}
    (E_{SF})_{i,j}= (\delta_{X'_i,Y'_j} ) \otimes \left[ \prod_{D_{SF}} O_{f_2,z} (E_{CCD})_{i,j} \right]
\end{equation}
where $z$ is the CCD–O2 distance (Fig. \ref{Fig_1}) and $O_{f, z}$
is the lens operator that transforms field $E_{CCD}$ at
lens distance $z$ into focal-plane field $E_{SF}$. The $\prod_{D_{SF}}$
two-dimensional gate operator accounts for the SF
aperture and defines the size of the $k$-space selected
zone. These operators work within the moving coordinates
of O2. Two-dimensional Dirac function
$\delta$ and convolution operator $\otimes$ perform the $X'_i,Y'_j$
displacement to yield $E_{SF}$ within the fixed coordinates
of O1. Operator $O_{f,z}$ can be expressed \cite{goodman2005introduction} as
\begin{eqnarray}\label{Eq_2}
 \nonumber   O_{f,z}&=& \int_{-\infty}^{-\infty}  dx \int_{-\infty}^{-\infty} dy\\
 && \times \exp[i 2\pi(x x' + y y')/\lambda f ] ~E(x,y)
\end{eqnarray}
where $O_{f,D}$ Fourier transforms $E(x, y)$, which is multiplied
by a quadratic phase function that corresponds
to $z - f$ propagation. To determine $E_{SF}$ over a wider
region it is necessary to combine the SF fields $(E_{SF})_{i, j}$
that correspond to various SF positions $i, j$:
\begin{eqnarray}\label{Eq_3}
 E_{SF}&=& \sum_{i}^{G} \sum_{j}^{G} (E_{SF})_{i, j}\\
\nonumber &=& \sum_{i}^{G} \sum_{j}^{G} (\delta_{X'_i,Y'_j})\otimes\left[\prod_{D_{SF}}O_{f_2,z} (E_{CCD})_{i, j} \right]
\end{eqnarray}
If the selected SF zones are contiguous and do not
overlap, $X'_{i+1}-X'_i= D_{SF}$ et  $Y'_{j+1}-Y'_j= D_{SF}$,
the wider region is $G$ times wider than for a single
hologram.

In our experiment, field $(E_{CCD})_{i, j}$ is measured in discrete
pixels, and the CCD measured pixels are inserted
into an $N \times N$ empty grid of step $d$. For fast Fourier
transforms (FFTs), $N$ is a factor of 2 larger than the
CCD pixel number. In the SF plane, field $(E_{SF})_{i, j}$ is
a FFT calculated on a grid of step $d'$. Inasmuch as
FFT steps $d$ and $d'$ obey $N d d' =\lambda f_2$, the SF grid's full
size is $N d'= D_{SF}$. The $\prod_{D_{SF}}$
 gate function is thus
accounted for by the FFT. To obtain the sum over
field $(E_{SF})_{i, j}$ that yields $E_{SF}$  [Eq. \ref{Eq_3}], we insert each
$N \times N$ small matrix $(E_{SF})_{i, j}$ within an $E_{SF}$ $N' \times  N'$
large matrix ($N'=G N$). Accounting for the $X'_i, Y'_j$
translation, the indices within the large matrix of the
center of the small matrix $i^{\textrm{th}}$, $j^{\textrm{th}}$ are $X'_i/d'$
 and $y'_j/d'$. Because the object is at a distance $z''$ of O1
(focal $f_1$), we calculate $E_{obj}$ by applying to $E_{SF}$ the
reverse lens operator $O^{-1}_{f_1, z''}$ whose expression can be
deduced from Eq. \ref{Eq_2}. Because $O^{-1}$ involves a FFT,
object plane grid step $d''$, which obeys $N' d' d'' = \lambda f_1$,
decreases linearly with $G$: $d'' = \lambda f/f_2 G)$. In measuring
CCD field $E_{CCD}$ that corresponds to different,
nonoverlapping positions of SF, one can thus calculate
object plane field $E_{obj}$ with a resolution and a pixel
size that improve linearly with the amount of data
acquired.

\begin{figure}
\begin{center}
  \includegraphics[width=8.5 cm]{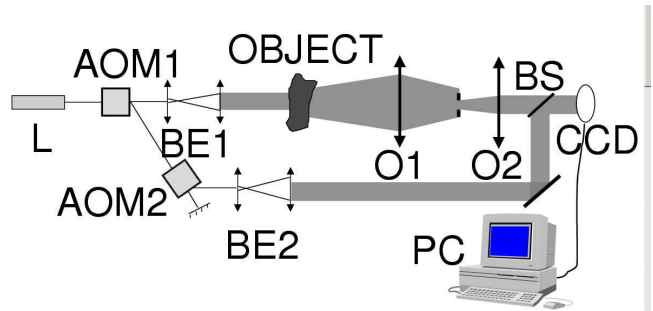}
  \caption{Heterodyne holography setup with laser L, accousto-
optic modulators AOM1 and AOM2, beam expanders
BE1 and BE2, objectives O1 and O2, beam splitter BS, a
CCD camera, and computer PC.}\label{Fig_2}
\end{center}
\end{figure}

Figure \ref{Fig_2} shows our experimental setup in details.
Laser L (633-nm He–Ne laser, or 850-nm laser diode)
is split into two beams (object and reference), which are
combined by a beam splitter (BS) in the CCD camera.
Two acousto-optic modulators (AOM1 and AOM2) are
used for shifting the reference beam by $\delta f = 6.25$ Hz
(25\% of the CCD image frequency). A frame grabber
and a Pentium II computer record the CCD-modulated
interference patterns and calculate complex field $E_{CCD}$
in the CCD. A step motors (0.1-$\mu$m step) allows O2,
the BS, the SF aperture ($D_{SF}=  1.84$, 1.9 mm in the
$x$ and $y$ directions) and the CCD, keeping them in
alignment.

The first object studied is a U.S. Air Force test target
lightened by a static speckle pattern emerging from
a ground-glass plate illuminated by the He–Ne laser.
In the CCD plane, the calculation step $d_x = 8.42 \mu$m
and $d_y = 8.3 \mu$m is equal to the pixel size. The matrix
dimension ($N \times N$, with $N = 1024$) is larger than
the CCD dimension $768 \times 576$. In the SF plane, the
grid step is $d'_x= 1.84 \mu$m and $d'_y= 1.86 \mu$m. The
CCD–O2 distance is $z = f_2 + 6.87$ cm, and the object–
O1 distance is $z'' \approx 5$ cm. To improve the resolution
by a factor $G = 4$ in the $x''$ direction we use
a synthetic-aperture grid dimension of $N' \times  N$, with
$N' =  4N = 4096$. In the object plane, the grid step
is then four times smaller in the $x''$ direction ($d''_x=d_x/4 = 4.21 \mu$m)  and is unchanged in the $y''$ direction
($d''_y = 16.6 \mu$m).

\begin{figure}
\begin{center}
  \includegraphics[width=8.5 cm]{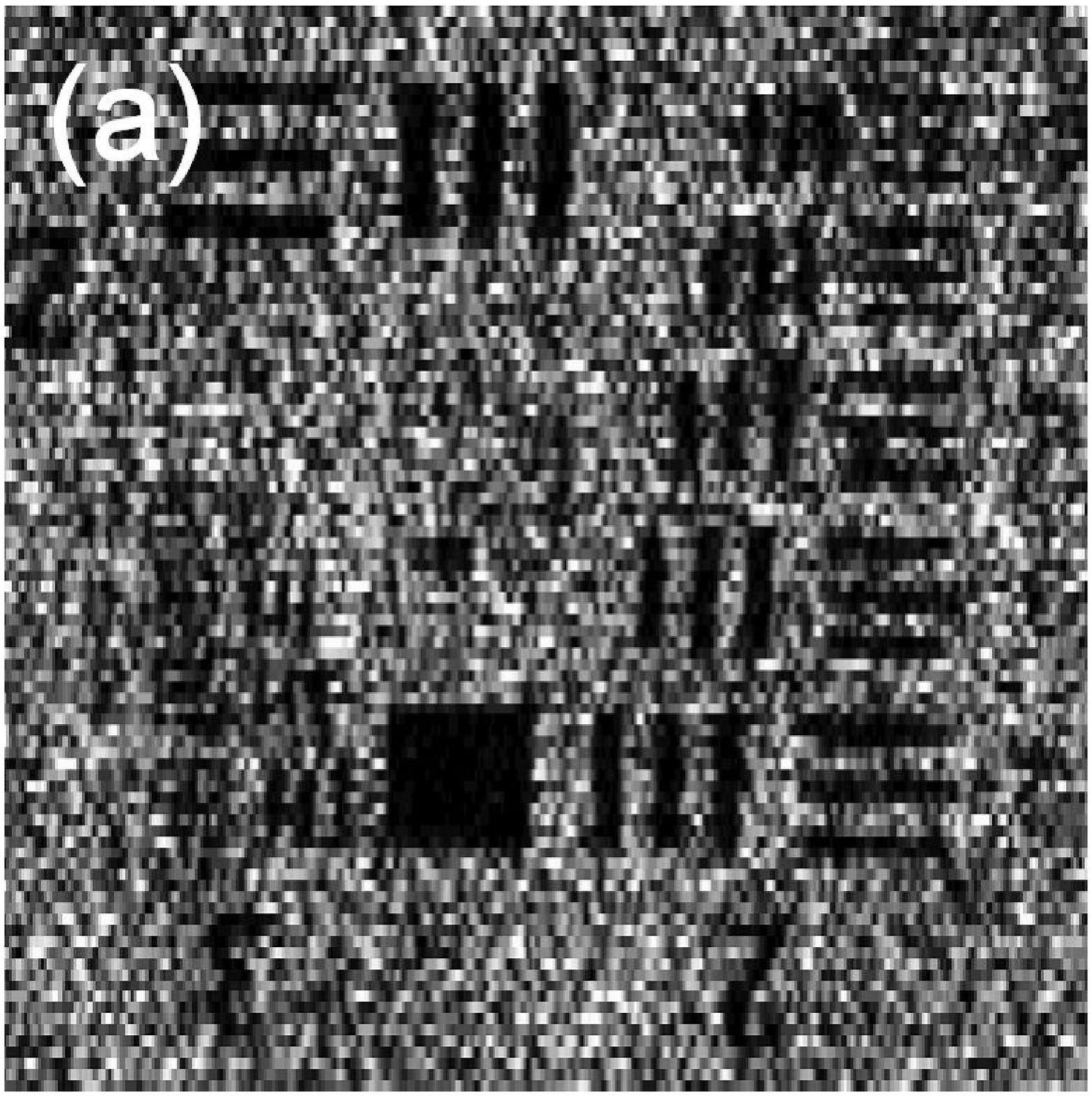}\\
 \includegraphics[width=8.5 cm]{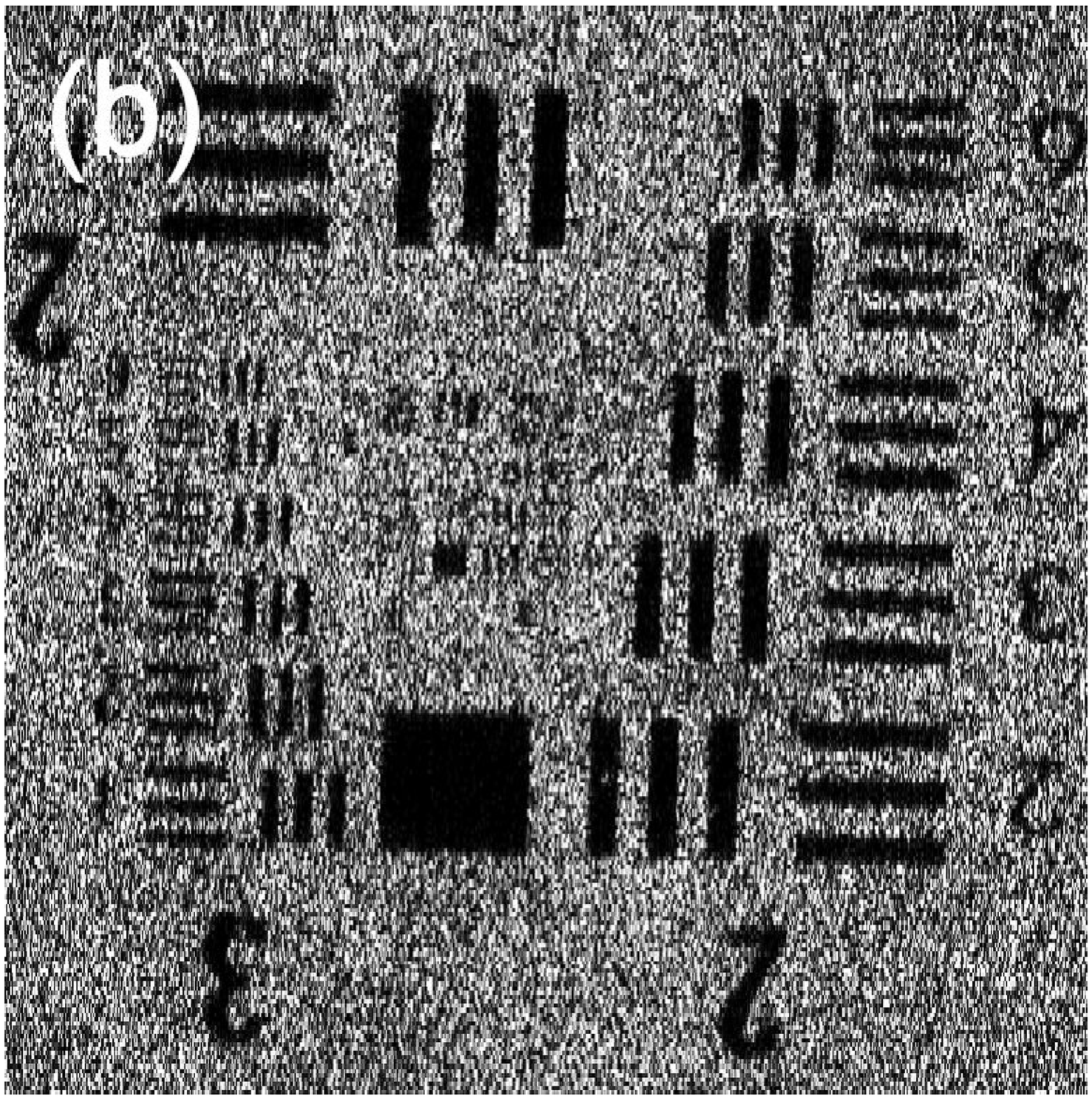}
  \caption{(a), (b), the center of the image of a U.S. Air Force
test target obtained without and with a synthetic aperture,
respectively. The reconstruction was performed
with $1024 \times 1024$ and $4096 \times 1024$ pixels, respectively; it
corresponds to a 17.2 mm $\times$ 17 mm image. The viewed
zone is 4.03 mm $\times$ 3.98 mm.}\label{Fig_3}
\end{center}
\end{figure}

We calculated for a single hologram the U.S. Air
Forcer intensity image ($N \times N = 1024 \times 1024$ pixels,
or 17.2 mm $\times$ 17 mm). The central part of the image
is shown in Fig. \ref{Fig_3}(a). The image resolution and
the speckle size, which depend on $k$-space extension
$D'= D_{SF}$ , should both equal $d''=(f_1/f_2)d$. In our
case, a low-pass filter, internal to the camera, lowers
the bandwidth of the analog video stream of data,
which degrades the resolution and enlarges the speckle
size by a factor of $\sim 2$ in the $x''$ direction.

When the synthetic-aperture algorithm is applied,
because the BS moves with SF the reference optical
path length changes for each SF position. Each measured
$E_{CCD}$ field map is thus shifted by an unknown
phase. To determine the phase correction we acquire
holograms such that the zones covered by the aperture
overlap for two consecutive SF positions $(X'_{i+1}-X'_i)<DSF$
($i = 1... 21$ and $X'_{i+1}-X'_i= 250 \mu$m in our
experiment). We assume thus that fields $(E_{SF})_{i+1,j}$
are equal in the overlapping region, and we
determine the phase by maximizing, in the overlapping
region, the $(E_{SF})_{i,j}$  to $(E_{SF})_{i+1,j}$  field correlation,
which we found to be 90\% at maximum. We calibrated
the $X'_{i+1}-X'_i$ transition (135 pixels) and the CCD–O2
distance [$z = 6.87$ mm in the quadratic phase factor of
Eq. \ref{Eq_2}] by this optimization process.

We calculated the synthetic-aperture reconstructed
image ($4096 \times 1024$ pixels, 17.2 mm $\times$ 17 mm) when
the SF, O2, the BS, and the CCD are $X'$ translated over
$D' = 5$ mm in 21 250-$\mu$m steps. The image center
is shown in Fig. \ref{Fig_3}(b). The motor calibration yields a
relative translation of 134.8 pixels, in good agreement
with the value obtained by correlation. As expected,
the synthetic-aperture image exhibits better resolution
and smaller speckle size. If the BS phase-shift correction
is not made, the speckle size, which depends on
the field extension in $k$-space $\sim G D_{SF}$, is the same,
but the resolution, which depends on the $k$-space field
coherence length ($\sim G D_{SF}$ with correction and $ \sim D_{SF}$
without), is lower. The image contrast is thus lower.

\begin{figure}
\begin{center}
  \includegraphics[width=8 cm]{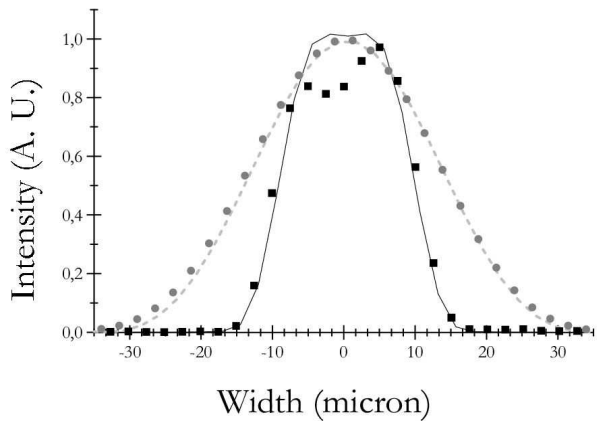}
  \caption{Cut in intensity for a 30-mm hole in the x and y directions.
Points, experimental synthetic-aperture reconstruction;
and curves, expected theoretical shapes. Dark
filled squares and the solid curve correspond to the $x''$ cut;
lighter filled circles and the dashed curve correspond to the
$y''$ cut (synthetic aperture is formed in the $x''$ direction).}\label{Fig_4}
\end{center}
\end{figure}

To perform a quantitative analysis of our synthetic-
aperture technique, we studied a narrow hole of
diameter $ \emptyset= 30$ mm located at $z = 5.5$ cm in front of
the SF. The hole is backilluminated by an 850-nm
laser diode. The SF, O2, the BS, and the CCD are
$X'$ translated over $D'= 5$ mm in 250-$\mu$m steps that
correspond to 101.6 pixels. O1 is removed, and two
Fourier transforms \cite{le2000numerical} propagate the field from the
SF plane to the object plane such that the grid size
( $d'_x= d''_x = 2.46 \mu$m, $d'_y= d''_y= 2.5 \mu$m is the
same in both planes. Figure \ref{Fig_4} presents, in the $x''$ and
$y''$ directions, the reconstructed intensities (points)
compared with the theoretical intensities (solid curve)
that account for pixel averaging and diffraction.
As expected, the edges are sharper in the $x''$ direction.
For both $x''$ and $y''$, the agreement with the
experimental points is excellent. This experiment,
performed without O1, can be interpreted as yielding
the synthetic aperture in real space for field $E_{SF}$
that we obtained by Fourier transforming ECCD. The
synthetic aperture here allows the image's angular
resolution to be increased to reach grid step limit d0
for the resolution on the image. One could also use
the real-space synthetic aperture in the CCD plane11
without transforming $E_{CCD}$ into $E_{SF}$, but the ultimate
resolution limit $d$ is much less $d'\ll d$.

The experiments presented here are examples of
ways in which synthetic apertures are obtained in
both real and Fourier space. In Fourier space, the
synthesis allows either the field of view in the far
field \cite{le2000numerical} (with respect to $N d^2/\lambda$) or the resolution for the
near field (U.S. Air Force experiment) to be improved.
In real space, the synthesis improves either the field
of view (near field) or the resolution (far field). In
the experiment with holes, the pixel size of field $E_{SF}$,
$d'$, is small, and the hole is pushed into the far field
$z''\gg N d'^2\lambda$. Note that for both the U.S. Air Force
experiment of Binet et al. \cite{binet2002short} and the hole experiment,
the synthesis yields better resolution, not the enlargement
of the field of view that could be obtained by a
simple scanning method.

We thank Thomson-CSF Optronique for its support
and J. Hare for help and fruitful discussions.

M. Gross's e-mail address is gross@lkb.ens.fr.


\end{document}